\documentclass[twocolumn]{aastex6}

\pdfoutput=1

\usepackage{amsmath}
\usepackage{amsfonts}
\usepackage{amssymb}
\usepackage{wasysym}
\usepackage{comment}

\begin{document}


\title{Relativistic Light Sails}


\author{
David Kipping\altaffilmark{1}
}

\affil{dkipping@astro.columbia.edu}

\altaffiltext{1}{Department of Astronomy, Columbia University, 550 W 120th St., New York, NY 10027}

\begin{abstract}
	
One proposed method for spacecraft to reach nearby stars is by accelerating
sails using either solar radiation pressure or directed energy.
This idea constitutes the thesis behind the \textit{Breakthrough Starshot}
project, which aims to accelerate a gram-mass spacecraft up to one-fifth
the speed of light towards Proxima Centauri. For such a case, the
combination of the sail's low mass and relativistic velocity render
previous treatments formally incorrect, including
that of Einstein himself in his seminal 1905 paper introducing special
relativity. To address this, we present formulae for a sail's acceleration,
first in response to a single photon and then extended to an ensemble. We show
how the sail's motion in response to an ensemble of incident photons is
equivalent to that of a single photon of energy equal to that of the ensemble.
We use this ``principle of ensemble equivalence'' for both perfect and
imperfect mirrors, enabling a simple analytic prediction of the sail's
velocity curve. Using our results and adopting putative parameters for
\textit{Starshot}, we estimate that previous relativistic treatments 
underestimate the spacecraft's terminal velocity by ${\sim}50$\,m/s for the same
incident energy, sufficient to miss a target by several Earth radii.
Additionally, we use a simple model to predict the sail's temperature and
diffraction beam losses during the laser firing period, allowing us to estimate
that for firing times of a few minutes and operating temperatures below
$300^{\circ}$C ($573$K), \textit{Starshot} will require a sail of which absorbs
less than 1 in 260,000 photons.

\end{abstract}




\section{Introduction}
\label{sec:intro}

One remarkable consequence of electromagnetism is that light carries
finite momentum \citep{maxwell:1865,compton:1923}.
Consequently, when light reflects off a surface, it imparts a small
momentum kick to the surface leading to radiation pressure, an effect
hypothesized about since at least the $17^{\mathrm{th}}$ century
\citep{kepler:1619}. Since the early $20^{\mathrm{th}}$ century, it has been
recognized that this effect could be utilized to propel spacecraft with large
mirror-like sails harvesting the momentum of incident 
photons\footnote{Tsiolkovsky \& Zander first discuss this possibility in 1925
as detailed in \citet{zander:1964}.}.
Whilst such sails were originally conceived with solar radiation in mind, the invention of 
lasers in the 1960s enables efficient laser sailing propulsion systems too 
\citep{marx:1966,redding:1967,forward:1984}.

Recently, the \textit{Breakthrough Starshot} project (simply \textit{Starshot}
in what follows) announced plans to develop the technology needed for a laser
sail nano-satellite capable of flying to the closest stars within a generation.
A proposed configuration is to fire an Earth-based array of gigawatt (or
greater) lasers onto a gram-mass, microchip-sized satellite which would be
accelerated up to approximately one-fifth the speed of light, reaching Proxima
Centauri in just over two decades (see \citealt{heller:2017} for deceleration
schemes).

Whilst a great deal of literature, experiments and even space flight
demonstrations of solar sailing exist \citep{kawaguchi:2008,mori:2009},
\textit{Starshot} is unique for two main reasons. First, the target speeds
are relativistic, and thus classical expressions suitable in the context of
solar sails become invalid. Second, \textit{Starshot} is designed to be
ultra-light, which means the mass of the sail cannot be assumed to be infinite,
as is typically assumed in relativistic calculations of photon exchanges with a
mirror (e.g. see \citealt{gjurchinovski:2013}).

In this work, we first present a simple derivation of the relativistic velocity
curve of a light sail in Section~\ref{sec:perfect}. We then extend our
analysis to consider the effect of imperfect mirrors and subsequent thermal
heating of the sail and spacecraft payload in Section~\ref{sec:imperfect}. We
finish with some key conclusions in Section~\ref{sec:discussion} and highlight
parts of the calculation requiring further work.

\clearpage

\section{Sailing with a Perfect Mirror}
\label{sec:perfect}

\subsection{A single photon}
\label{sub:single}

\begin{figure*}
\begin{center}
\includegraphics[width=16.0cm,angle=0,clip=true]{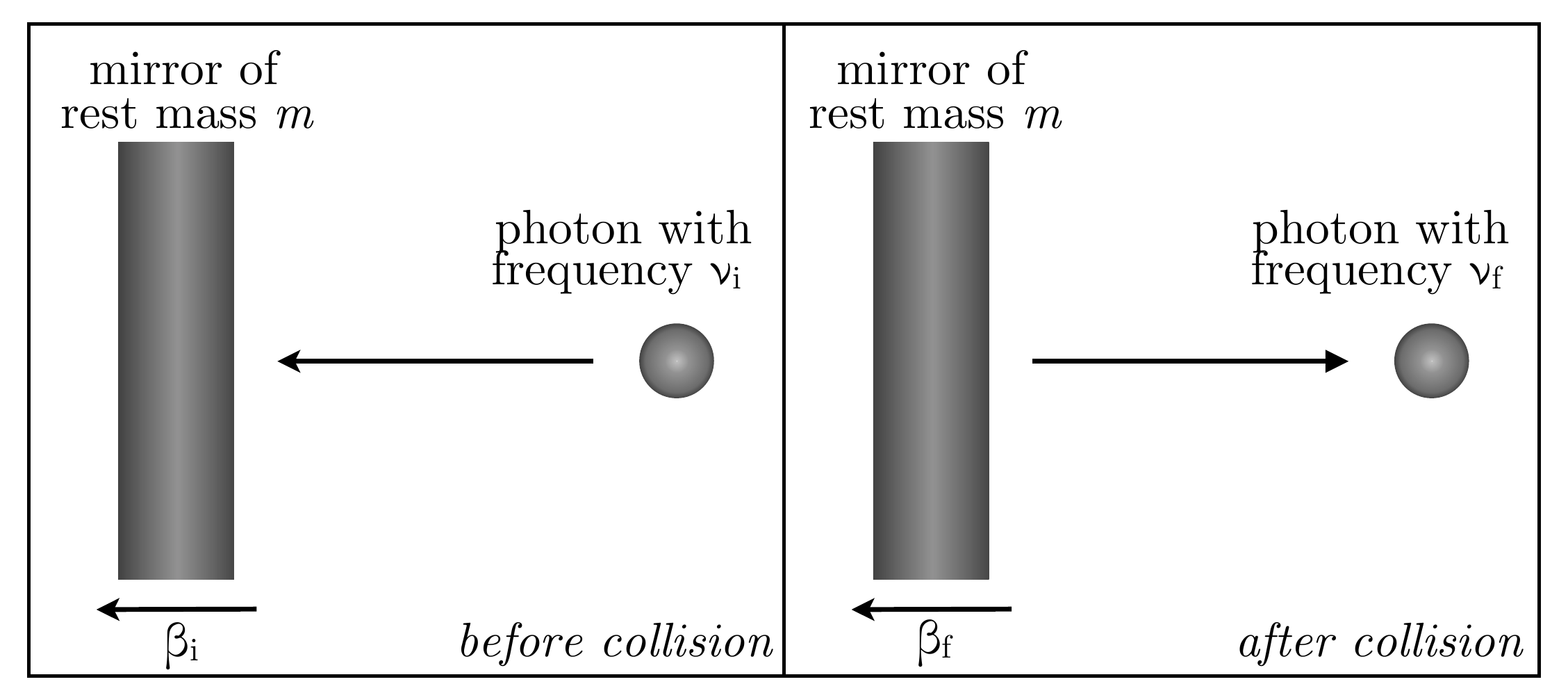}
\caption{\emph{
Schematic depicting the before and after configurations of a photon and a mirror.
}}
\label{fig:figure1}
\end{center}
\end{figure*}

We begin by considering the simple case of a single photon of frequency $\nu_i$
fired at a normal incident angle towards a perfect mirror, or equivalently a
light sail, of mass $m$ moving along the same vector as the photon at speed
$c \beta_i$, as depicted in Figure~\ref{fig:figure1}. The reflection of the
photon is assumed to be perfectly elastic, although we later relax this
assumption in Section~\ref{sec:imperfect}. The motion of the mirror and the
frequency of the photon can be calculated by requiring the conservation of
relativistic energy and momentum. 

First, the system's total energy (the sum of the photon's energy and the
mirror's energy) must be conserved before and after the reflection. Using the
relativistic expressions, one may write that

\begin{align}
h \nu_i + \frac{m c^2}{\sqrt{1-\beta_i^2}} &= h \nu_f + \frac{m c^2}{\sqrt{1-\beta_f^2}}.
\label{eqn:energy}
\end{align}

Similarly, requiring that the system conserves momentum, we find

\begin{align}
\frac{h \nu_i}{c} + \frac{m \beta_i c}{\sqrt{1-\beta_i^2}} &= -\frac{h \nu_f}{c} + \frac{m c \beta_f}{\sqrt{1-\beta_f^2}}.
\label{eqn:momentum}
\end{align}

Solving the Equations~(\ref{eqn:energy}) \& (\ref{eqn:momentum}) simultaneously
and simplifying, we find

\begin{align}
\beta_f &= \frac{
4r^4 (1-\beta_i)^2 + \beta_i + 2r(1-\beta_i)\sqrt{1-\beta_i^2} - 2r^2(1-\beta_i^2)
}{
1 - 4 r^2 (1-\beta_i) (\beta_i - r^2(1-\beta_i))
},
\label{eqn:betaf}
\end{align}

and

\begin{align}
\nu_f &= \nu_i \Bigg( \frac{1-\beta_i}{ 1 + \beta_i + 2 r \sqrt{1-\beta_i^2} } \Bigg),
\label{eqn:nuf}
\end{align}

where we have defined $r$ as the photon's ``relative energy'' using

\begin{align}
r &= \frac{h \nu_i}{m c^2}.
\end{align}

Note, that our calculation has ignored the effect of the mirror's gravity,
which in principal imparts a small gravitational frequency shift which is
negligible for gram-mass sails.

\subsection{Redshift of the reflected photon}
\label{sub:redshift}

\begin{figure}
\begin{center}
\includegraphics[width=8.4cm,angle=0,clip=true]{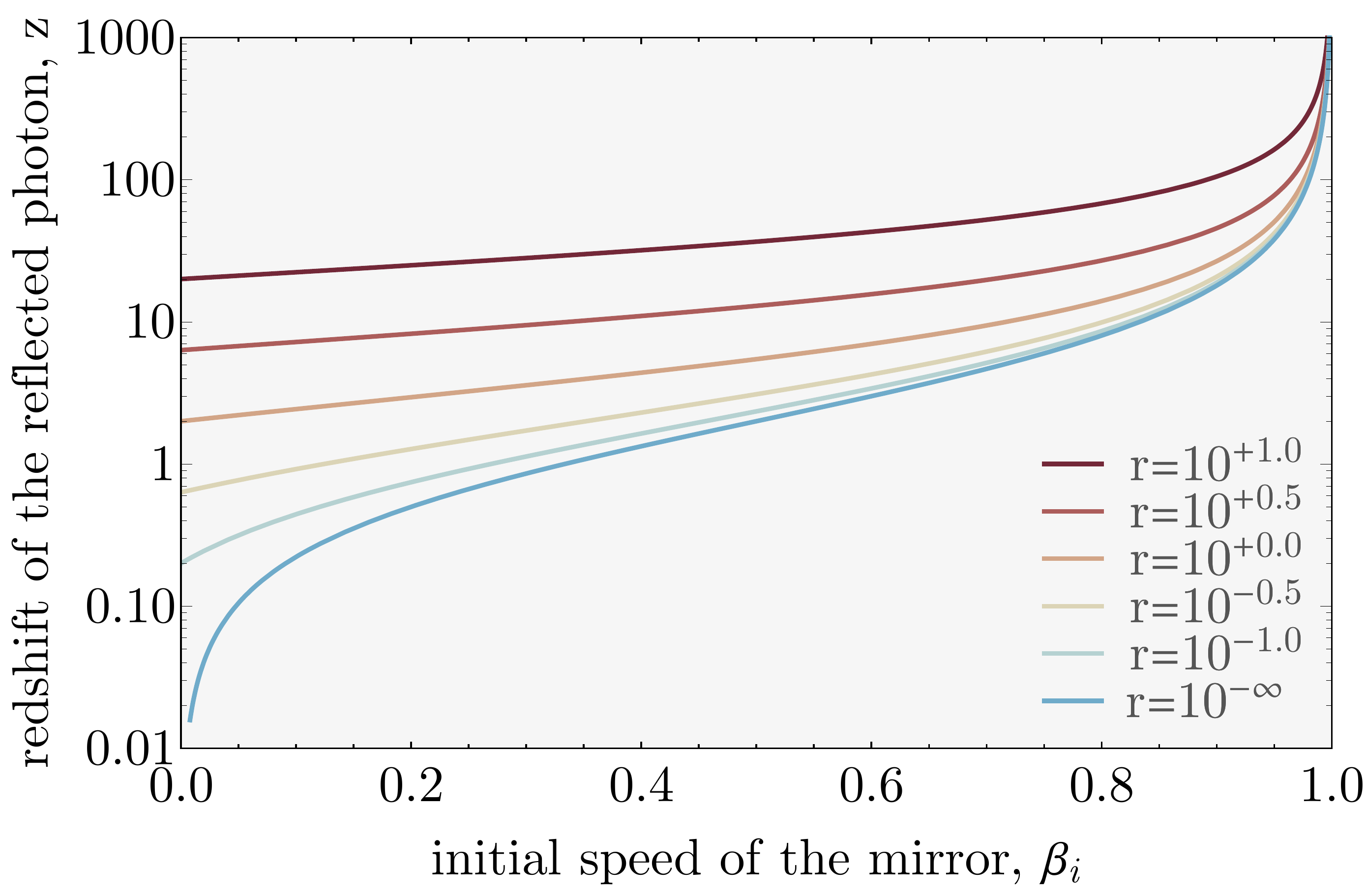}
\caption{\emph{
Redshift of a photon reflected off a mirror moving at relativistic
speed along the momentum vector of the incident photon. 
}}
\label{fig:redshift}
\end{center}
\end{figure}

Equation~(\ref{eqn:nuf}) may also be expressed in terms of the redshift, $z$, of
the reflected photon:

\begin{align}
1 + z &= \frac{ 1 + \beta_i + 2 r \sqrt{1-\beta_i^2} }{1-\beta_i},
\label{eqn:redshift}
\end{align}

which we plot in Figure~\ref{fig:redshift} for several choices of $r$.
Equation~(\ref{eqn:redshift}) reveals that the reflected photon will have a
redshift of zero when

\begin{align}
\beta_{i,z=0} &= -\frac{r}{\sqrt{1+r^2}},
\end{align}

where the negative sign indicates that the mirror is now coming towards
the photon.

We also point out that Equation~(\ref{eqn:redshift}) and
Figure~\ref{fig:redshift} reveal that the photon becomes
redshifted to infinity (i.e. redshifted out of existence)
as $\beta_i\to1$. This result implies that as the mirror
moves closer to $c$, the transfer of the photon's energy
into the kinetic energy of the mirror becomes increasingly
efficient. This result is verified later in
Section~\ref{sub:efficiency}.

\subsection{Accelerating a mirror with a single photon}

In the limit of $\beta_i\to0$, in other words an initially stationary mirror, one may
write that the mirror will be accelerated up to a speed of

\begin{align}
\lim_{\beta_i\to0} \beta_f &= \frac{2 r (1+r)}{1+2 r (1+r)}.
\label{eqn:betaf0}
\end{align}

Solving the above for $\beta_f=\tfrac{1}{2}$ yields a characteristic relative
photon energy, $r$, necessary to impart relativistic motion as

\begin{align}
r_{\mathrm{rel}} &= \frac{\sqrt{3}-1}{2} \simeq 0.366...
\label{eqn:characteristic}
\end{align}

To first order in $r$, Equation~(\ref{eqn:betaf0}) is simply
$2r$, which shows that $r \gtrsim \mathcal{O}[10^{-2}]$
to get to even a few percent the speed of light.

\subsection{Efficiency}
\label{sub:efficiency}

In the case of photon sailing, the primary goal of hitting the sail with photons
is to propel a sail in the desired direction. Two useful figures of merit to
consider in this context are the kinetic energy and speed of the sail in response
to a photon reflection.

Consider first: what is the gain in kinetic energy of the mirror as a function
of its initial velocity, $\beta_i$? The change in kinetic energy of the mirror
is most easily expressed by equating it to the total energy lost by the
photon:

\begin{align}
\Delta E_K &= h \nu_i - h \nu_f,\nonumber\\
\epsilon_K \equiv \frac{\Delta E_K}{h\nu_i} &= 1 - \frac{\nu_f}{\nu_i}.
\end{align}

where on the second line we have re-expressed the kinetic energy
gain in units of the incident photon's energy, which can be considered
to be the efficiency by which energy is transferred from the photon
to the sail. Using Equation~(\ref{eqn:nuf}), we may now write that

\begin{align}
\epsilon_K &= 1 - \frac{1 - \beta_i}{1+\beta_i+2r \sqrt{1-\beta_i^2}}.
\end{align}

In the limit of $r\to0$, where the photon's energy is much less than the
rest mass energy of the sail, we find that

\begin{align}
\lim_{r\to0} \epsilon_K &= \frac{ 2\beta_i}{1+\beta_i},
\end{align}

which is a monotonically \textit{increasing} function from $\beta_i=0\to1$.
This result therefore demands that the fraction of the photon's energy
transferred to the sail as kinetic energy increases as $\beta_i$ increases.
In this sense, the sail becomes more efficient
once it has gained some initial momentum, verifying the argument based earlier
in Section~\ref{sub:redshift}.

Consider now the velocity change of the sail
as a function of $\beta_i$. In the classical framework, the speed
ever-increases linearly into the super-luminal regime. At low velocities,
one may easily show that our expression in Equation~(\ref{eqn:betaf})
reproduces the classical behavior; for example, in the limit of
$r\ll1$, the related expression Equation~(\ref{eqn:betaf0}) simply gives
$2r$, as expected. Therefore, the relative velocity change predicted by
our formula should decrease at high speeds, in order to reproduce
an asymptote towards $c$. One can verify this mathematically by
writing

\begin{align}
\beta_f - \beta_i &= \Big(2 (1-\beta_i) \sqrt{1-\beta_i^2}\Big) r + \mathcal{O}[r^2],
\end{align}

which reproduces the correct behavior of a velocity change of $2 r$
at low $\beta_i$ and zero velocity change as $\beta_i$ approaches
unity.

\subsection{Accelerating a mirror with an ensemble of photons}
\label{sub:ensemble}

We now consider firing multiple photons at the sail/mirror in order to induce
acceleration. In what follows, we ignore the effect of drag forces, such
as interstellar dust and even photonic gas drag \citep{balasanyan:2009}.
We treat each photon as striking the mirror consecutively, leading to a series
of small impulses, each of which increases the velocity of the sail
slightly.

We set the initial velocity to $\beta_0=0$ and then define $\beta_n$, the
velocity after the $n^{\mathrm{th}}$ reflection, as $\beta_f$ from
Equation~(\ref{eqn:betaf}), replacing $\beta_i \to \beta_{n-1}$. Writing
out the first few terms and simplifying, one may show that $\beta_n$ may
be written as

\begin{align}
\beta_n &= 1 - \frac{1}{1+2nr(1+nr)}.
\label{eqn:betan}
\end{align}

Comparing this expression to $\beta_f$ in Equation~(\ref{eqn:betaf}),
one can see that

\begin{align}
\beta_n &\simeq \lim_{\beta_i\to0} \beta_f \Big|_{r \to n r}.
\label{eqn:conject}
\end{align}

In other words, the final speed of the light sail after $n$
reflections of identical photons of relative energy $r$ is equal to
that expected due to the reflection of single photon of energy $n r$.
We refer to this as the principle of ensemble equivalance in the
remainder of this work.

\begin{figure}
\begin{center}
\includegraphics[width=8.4cm,angle=0,clip=true]{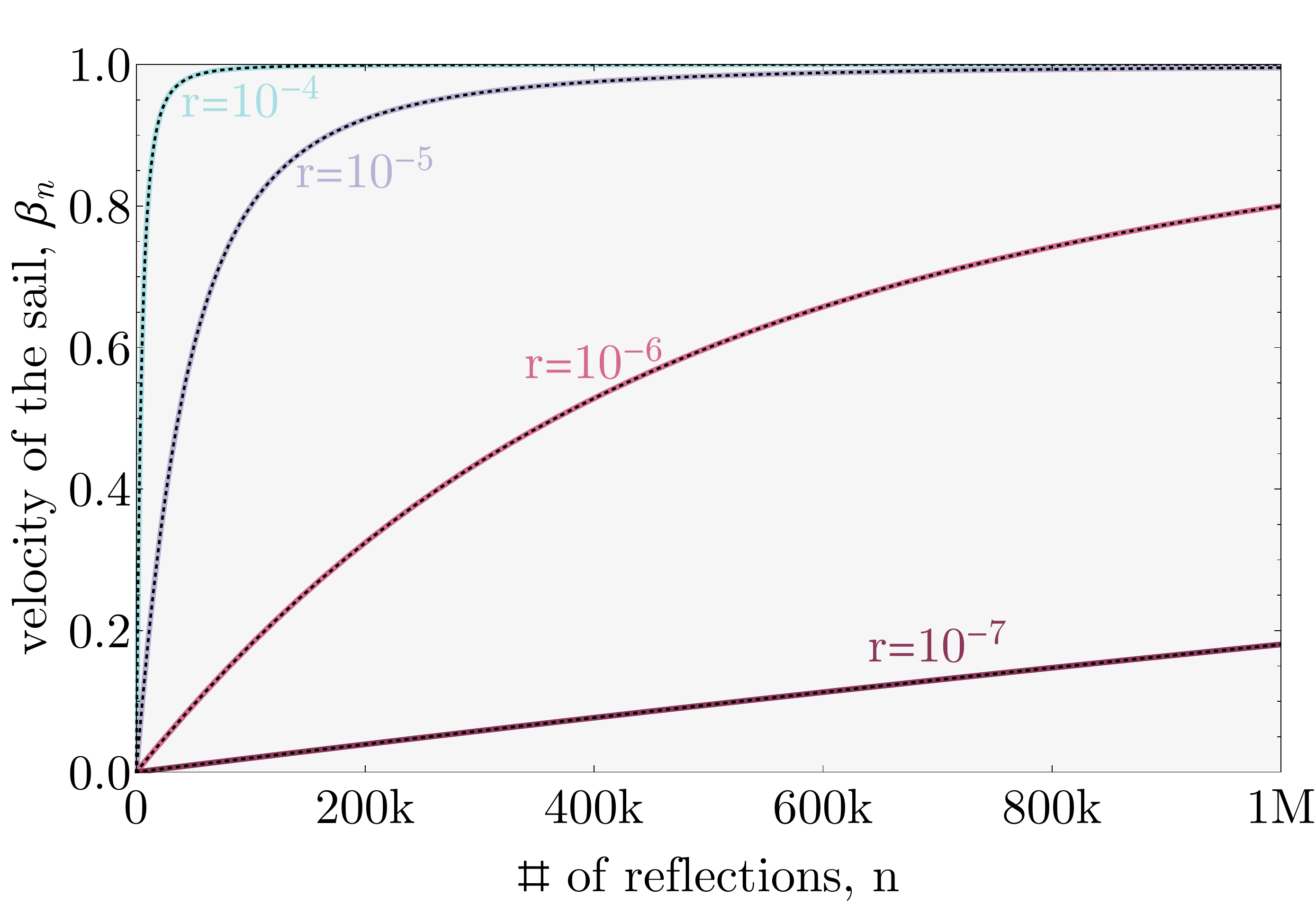}
\caption{\emph{
Velocity curves as a function of number of reflected photons onto
the sail. The solid lines are from numerically iterating on
Equation~(\ref{eqn:betaf}), whereas the dashed lines are from the simple
analytic equation derived using our conjecture. Each set of lines
is for unique choices of $r$, the ratio of the incident photon energy
to the sail's rest mass.
}}
\label{fig:curves}
\end{center}
\end{figure}

In Figure~\ref{fig:curves}, we compare the velocity curve predicted by
Equation~(\ref{eqn:conject}) versus that computed numerically for
$10^6$ reflections for large choices of $r$. These experiments find the
formulae are correct to within machine precision, thereby providing a
simple formula to predict the velocity curves of relativistic sails.

By re-arranging Equation~(\ref{eqn:betan}) to make $n$ the subject, we
are able to write down a simple formula for the number of photons needed
to accelerate a sail up to a target relativistic speed,
$\beta_{\mathrm{targ}}$:

\begin{align}
n &= \frac{1}{2 r} \Bigg( \sqrt{ \frac{1+\beta_{\mathrm{targ}}}{1-\beta_{\mathrm{targ}}} } - 1 \Bigg),
\label{eqn:nreq}
\end{align}

or, equivalently, that the total light energy needed to strike the
mirror is

\begin{align}
E_{\mathrm{light}} &= \tfrac{1}{2} m c^2 \Bigg( \sqrt{ \frac{1+\beta_{\mathrm{targ}}}{1-\beta_{\mathrm{targ}}} } - 1 \Bigg).
\label{eqn:Ereq}
\end{align}

\begin{figure*}
\begin{center}
\includegraphics[width=17.0cm,angle=0,clip=true]{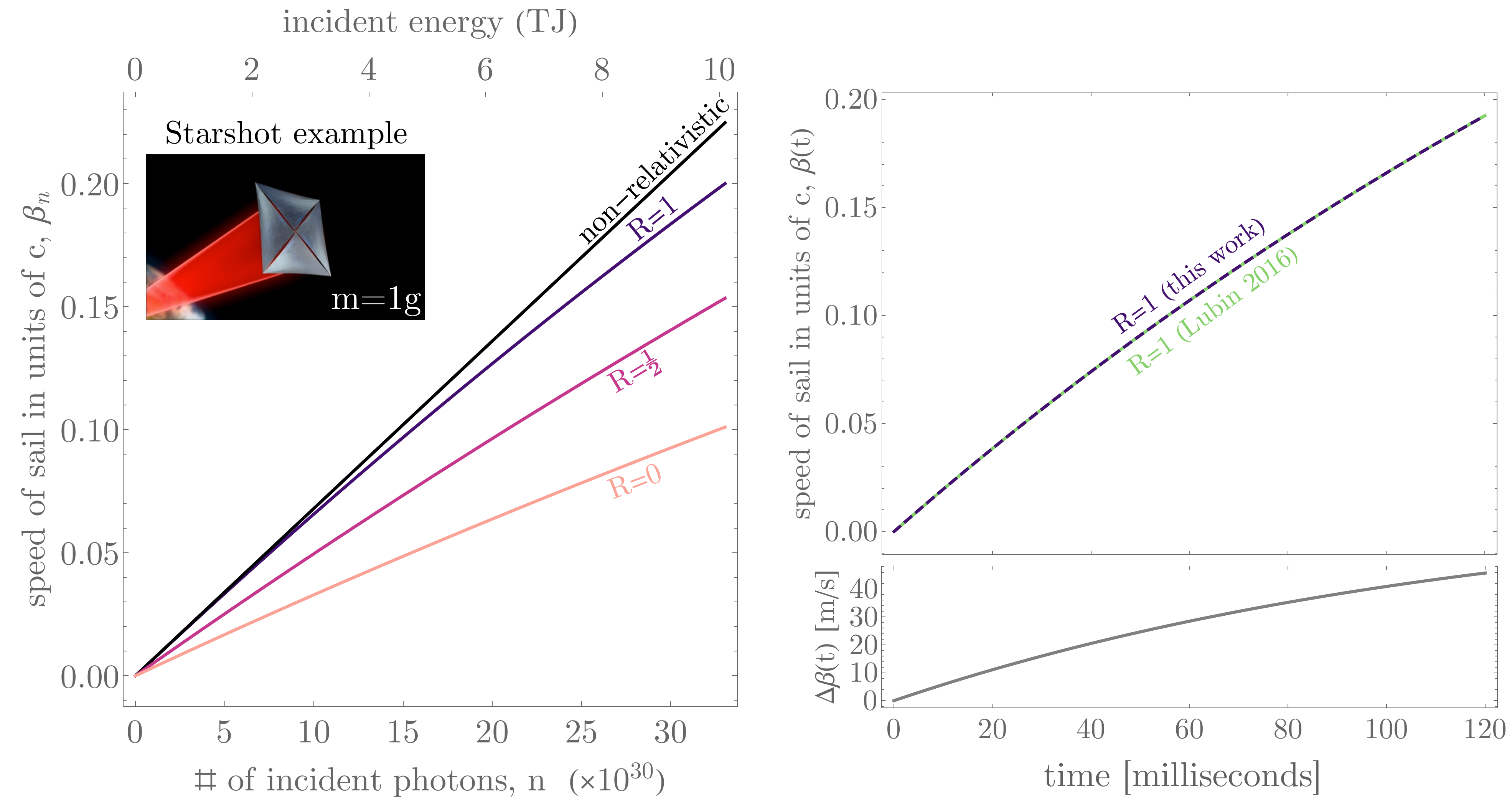}
\caption{\emph{
\textbf{Left:} Velocity curve for a Starshot-like choice of parameters comparing
the difference between the predictions of different formulae. The
laser is assumed to be at $\lambda=650$\,nm wavelength and the sail
mass is $m=1$\,g. The different colored lines are for different
choices of $R$, the sail's reflection coefficient.
\textbf{Right:} To compare to the literature formula of \citet{lubin:2016}
and \citet{kulkarni:2016}, we replace reflection number, $n$, with time.
Formulae show close agreement although slight residuals (lower right) are
present as a result of the infinite sail mass assumption of \citet{lubin:2016}.
}}
\label{fig:starshot}
\end{center}
\end{figure*}

As a practical example, we plot the velocity curve of a \textit{Starshot}-like
sail ($m=1$\,g) accelerating up to $0.2c$ using our relativistic 
formula in comparison to the non-relativistic case in
Figure~\ref{fig:starshot}.

\subsection{Previous literature \& why Einstein's formalism is erroneous for Starshot}
\label{sub:comp}

It is instructive to compare our results to those of the pre-exisiting
literature. Our solution calculates two distinct quantities: the
redshift of a single photon after reflection (Equation~\ref{eqn:nuf})
and the resulting velocity change of the sail (Equation~\ref{eqn:betaf}),
which forms the basis to scale up to an ensemble of photons
(Equation~\ref{eqn:betan}).

A photon's redshift off a relativistic mirror is a classic problem which has
been studied by many previous authors, including Einstein himself in his
historic paper introducing special relativity \citep{einstein:1905}. The
corresponding velocity change of the mirror is less commonly derived,
although our derivation finds that the solutions must come as a pair. Of course
for beamed laser sailing, it is this velocity change which is of greatest
interest. Before comparing our velocity predictions to the literature,
we first consider the redshift result, due to the rich literature of 
comparisons at our disposal.

We first compare to \citet{gjurchinovski:2013} who provide a pedagogical
derivation of a photon incident upon a relativistic mirror at an angle
$\alpha$ but under the explicit assumption of an infinitely heavy mirror
($m\to\infty$). By conserving energy and momentum,
\citet{gjurchinovski:2013} obtain

\begin{align}
\nu_f &= \nu_i \Bigg( \frac{1 - 2\beta_i\cos\alpha + \beta_i^2 }{1-\beta_i^2}\Bigg).
\label{eqn:gjurchinovski}
\end{align}

As expected, the above is equivalent to our Equation~(\ref{eqn:nuf}) in the
case of a normal incident photon ($\alpha=0$), as was assumed in our work, and
the limit of $r\to0$ (which is equivalent to \citeauthor{gjurchinovski:2013}'s
assumption of $m\to\infty$).

Another insightful example to compare to (where it cannot be assumed that
$m\to0$) is for \citeauthor{compton:1923} scattering, which is essentially the
same problem but where the mirror is replaced with an electron. For an electron
initially at rest ($\beta_i=0$), the photon's frequency is shifted to
(Equation 7.2 of \citealt{rybicki:1979}):

\begin{align}
\nu_f &= \frac{\nu_i}{1 + r (1-\cos\theta)},
\label{eqn:compton}
\end{align}

where $\theta$ is the scattering angle equal to $\pi$ for an exact
reflection back along the original path - as adopted in our work.
As expected, Equation~(\ref{eqn:compton}) is indeed equivalent to our result
in Equation~(\ref{eqn:nuf}) for $\beta_i=0$ and $\theta=\pi$.

Having established the validity of our redshift formula, we now compare
it to that being used in the literature of light sails. Of most relevance
is the result curated in the ``Roadmap to Interstellar Flight'', a
comprehensive review by \citet{lubin:2016}, which ultimately inspired the
\textit{Breakthrough Starshot} project \citep{popkin:2017}. \citet{lubin:2016}
report that their relativistic solutions come from \citet{kulkarni:2016}, who in 
Equation~(1) have

\begin{align}
\lambda_f &= \lambda_i \gamma^2 (1+\beta)^2,\nonumber\\
\qquad&= \lambda_i \Bigg(\frac{1-\beta_i}{1+\beta_i}\Bigg),
\end{align}

which is equivalent to \citet{gjurchinovski:2013} for $\alpha=0$
and also to our Equation~(\ref{eqn:nuf}) in the limit of $r\to0$.
Therefore, although it is not explicitly stated in \citet{kulkarni:2016},
the authors appear to have tacitly adopted the infinite mass sail
approximation\footnote{
In a subsequent paper by \citet{kulkarni:2017}, it
is explicitly verified that this is indeed an assumption made in their
derivation (see Section 2 of that work).
}.

This can be verified by following the description of their derivation,
which unlike this work and \citet{gjurchinovski:2013} uses Lorentz frame
transfers rather than balancing conserved quantities. Specifically, the
authors first shift the photon to the sail's frame, then assume it
``is emitted with the same wavelength as it is incident with'', before
finally transferring back to the original frame. Crucially, this is also
the same tacit assumption made by Einstein himself in \citet{einstein:1905},
who in Section~8 of that work adopt same derivation procedure of frame
transfers, and use the same intermediate step in the sail's rest frame of
$\nu''=\nu'$ in Einstein's original notation (which indicates that the
reflected light's frequency equals the incident light's frequency when viewed
in the sail's frame).

It is with some trepidation that we argue here that Einstein, and indeed all
subsequent authors adopting this assumption (e.g.
\citealt{gjurchinovski:2013,lightman:1975,galli:2012}), must be formally wrong.
For a sail (or mirror) at rest, the reflected photon cannot have the same
frequency as the incident photon without violating the conservation of energy.
The photon has reversed momentum and so the mirror must increase its absolute
momentum (from initially zero, since it is defined to be at rest) to conserve
total momentum. Since the mirror is now moving, its kinetic energy must have
also increased. Therefore, to conserve the total energy of the system, the
photon has to lose energy which it can only do so by decreasing in frequency.
Ergo, Einstein's assumption that $\nu''=\nu'$ violates the conservation of
energy (note that this can also be seen by comparison to Compton scattering
where this general statement is false; \citealt{rybicki:1979}).

Another way to think about the above is to assume Einstein is correct
and that $\nu''=\nu'$ and then look at the consequences. The equality
$\nu''=\nu'$ means that the photon has lost no energy when reflecting off
a mirror at rest (since $E=h\nu$). If this is true, then by conservation
of energy, the mirror cannot have gained any kinetic energy. In other
words, the mirror does not move. This simple point demonstrates the
falsehood of $\nu''=\nu'$, since it requies that no matter how many photons
are incident upon a mirror initially at rest, it will never move. In other
words, $\nu''=\nu'$ would make the entire concept of light sailing
impossible, since objects could never be accelerated away from being
initially at rest.

Although formally wrong, one might argue that practically speaking this
infinite mass mirror assumption is always extremely well justified. In
other words, one might reasonably posit that whether this assumption is
imposed or not, the resulting predictions will be nearly identical.
Remarkably, this appears to be false. Consider the other half of the
solution now, the corresponding velocity change of the mirror in
response to an ensemble of photons (which we state in
Equation~\ref{eqn:betan}). This solution does not appear in
\citet{einstein:1905} but is derived in \citet{kulkarni:2016}, who,
recall used the same derivation framework for $\nu$ as Einstein.

\citet{kulkarni:2016} relate the relativistic velocity of a perfect sail in
response to a constant beam of power $P$ fully on the sail for a time $t$
as

\begin{align}
\tilde{t} &= \frac{1}{6} \Bigg[ \frac{(1+\beta_f)(2-\beta_f)}{(1-\beta_f)\sqrt{1-\beta_f^2}} - 2 \Bigg],
\label{eqn:tdash}
\end{align}

where we use $\tilde{t}\equiv (P t)/(m c^2)$. Although it was not stated in
\citet{lubin:2016} or \citet{kulkarni:2016}, we may re-arrange
Equation~(\ref{eqn:tdash}) to solve for $\beta$, which leads to a cubic
equation with one real root of

\begin{align}
\beta_f =& 1 - \frac{(\kappa - 2 - 6 \tilde{t})^{1/3}}{\kappa} \nonumber\\
\qquad& - \frac{(1+\sqrt{3}\mathrm{i})/2}{[\kappa^2(-5+2\kappa+6\tilde{t}(\kappa-4-6\tilde{t}))]^{1/3}},
\end{align}

where $\kappa\equiv\sqrt{5+12\tilde{t}(2+3\tilde{t})}$. Our work does not strictly assume
a constant laser illumination, which we would argue is an advantage of our
prescription, but it can be modified to such a form as follows. In the original
version of this manuscript, we accomplished this by taking
Equation~(\ref{eqn:betan}) and replacing $n r = E_{\mathrm{inc}}/(mc ^2) =
(P t)/(mc^2)=\tilde{t}$. In a reply that version, \citet{kulkarni:2017} correctly
point out that this does not account for the time delay for light to reach the
sail, leading to an unfair comparison of the two formulae and we correct for
this here. One may show that the time of the $n^{\mathrm{th}}$ photon reflection
on the sail, accounting for time delays, is given by

\begin{align}
t_n &= t_0 + \delta t \sum_{j=1}^n (1-\beta_{j-1})^{-1},
\end{align}

where $t_0$ is a reference time and $\delta t$ represents the time between
each photon emission (assuming a uniform rate i.e. constant power). Assuming
that the sail begins from rest at time $t_0$, we may use
Equation~(\ref{eqn:betan}) to write that

\begin{align}
t_n &= t_0 + \delta t \Big( n + n r (n - 1) + n r^2 (\tfrac{1}{3} - n + \tfrac{2}{3} n^2) \Big).
\label{eqn:timelag}
\end{align}

We may now re-arrange the above to make $n$ the subject and replace the real root
of the resulting cubic into Equation~(\ref{eqn:betan}) to give us a formulae
for $\beta$ as a function of time under constant laser power. Our formulae,
written as a function of time, is compared directly to that of \citet{lubin:2016}
and \citet{kulkarni:2016} in the right panel of Figure~\ref{fig:starshot}.
Although the two equations show close agreement for the fiducial choice of
parameters in Figure~\ref{fig:starshot}, they are not equivalent - as evident from
the residual plot in that figure. Specifically, our formula predicts a slightly
faster acceleration, due to the additional recoil accounted for by the photon
reflections ignored in the \citet{lubin:2016} formalism.

Although both formulae are fairly unwieldy when expressed in terms of time,
we can take the difference between them (the residuals) and perform a
series expansion in $r$. This leads to the following expression for the
difference between the two

\begin{align}
\beta[\mathrm{this\,\,work}] - \beta[\mathrm{Lubin\,\,2016}] &= 2 r \Big(
\tilde{t} - 4 \tilde{t}^2 + \tfrac{28}{3} \tilde{t}^3 + \mathcal{O}[\tilde{t}^4]
\Big)
\end{align}

For a target speed of $0.2c$, this corresponds to a difference of 47\,m/s,
which would change the arrival time at Proxima Centauri b by 8.7\,minutes.
Given planet b's orbital velocity, this would cause the planet to be in a
different location by 25,000\,km, or around four planetary radii.
Although the difference is certainly small, we highlight several key advantages
of this work's formalism of that of \citet{lubin:2016} and \citet{kulkarni:2016}:

\begin{itemize}
\item The formalism of \citet{lubin:2016} and \citet{kulkarni:2016} is predicated
on the assumption of no photon recoil on the sail, which technically makes it
impossible to ever accelerate the sail from rest.
\item It is not necessary to assume constant power on the sail with our formalism,
any temporal profile (for example one accounting for diffraction) can be employed.
\item If the arrival position of a relativistic sail to a nearby star needs
to be predicted to a precision of several Earth radii or better (for example if
attempting a fly-by manoeuvre), then our formula would be favored.
\end{itemize}

\section{Imperfect Sails in Thermal Equilibrium}
\label{sec:imperfect}
 
\subsection{Overview}

Throughout Section~\ref{sec:perfect}, we have explicitly assumed a perfect
mirror, one with a reflection coefficient of unity. In such a case, the
sail is maximally efficient and thermally stable, absorbing no photons as
thermal energy. Accordingly, the time frame over which one fires the
photons at the sail is inconsequential, and in principle, the sail can
receive the full jolt of energy in a single laser pulse. In practice,
even slight imperfections in the reflectivity will both degrade the
rate of acceleration and lead to the sail absorbing thermal photons,
potentially leading to a catastrophic failure of the sail and/or
electronic payload. We here provide a simple derivation of the
magnitude of these effects, starting again from the case of a single
photon.

\subsection{Inelastic photon collisions}
\label{sub:inelastic}

We begin by considering a single photon which makes an inelastic collision
with the mirror. The picture is therefore the similar to that depicted
in Figure~\ref{fig:figure1}, except the final photon is not reflected but
absorbed into the mirror, slightly increasing the rest mass energy of the
mirror. As before, we proceed by balancing the energy

\begin{align}
h \nu_i + \frac{m c^2}{\sqrt{1-\beta_i^2}} &= \frac{\mathcal{M} m c^2}{\sqrt{1-\beta_f^2}},
\end{align}

and momentum

\begin{align}
\frac{h \nu_i}{c} + \frac{m \beta_i c}{\sqrt{1-\beta_i^2}} &= \frac{\mathcal{M} m c \beta_f}{\sqrt{1-\beta_f^2}},
\end{align}

in the system, which may be solved for $\beta_f$ and $\mathcal{M}$, where
$\mathcal{M}$ is the relative increase in the rest mass energy of the sail, giving

\begin{align}
\mathcal{M} &= \Bigg( \frac{r(1-\beta_i^2)+\sqrt{1-\beta_i^2}}{ \sqrt{(1-r^2(1-\beta_i^2))^2} } \Bigg) \Bigg( \frac{\sqrt{1-\beta_i}}{1-\beta_i^2}\Bigg) \Bigg( 1+\beta_i \nonumber\\
\qquad& - 2 r \beta_i \sqrt{1-\beta_i^2} - r^2 (3-\beta_i) (1-\beta_i) (1+\beta_i) \nonumber\\
\qquad& + 2r^3(1-\beta_i^2)^{3/2} \Bigg)^{1/2},
\end{align}

and

\begin{align}
\beta_f^{\mathrm{abs}} &= \frac{\beta_i + r (1-\beta_i) \sqrt{1-\beta_i^2} - r^2 (1-\beta_i^2)}{1 - r^2(1-\beta_i^2)}.
\label{eqn:betaf_abs}
\end{align}

Note that we now distinguish between the mirror's velocity from
an absorbed versus reflected photon using the superscripts
``abs'' and ``ref'', respectively.
Accordingly, comparing Equations~(\ref{eqn:betaf}) \&
(\ref{eqn:betaf_abs}), we can verify the classical result that

\begin{align}
\lim_{r \ll 1} \lim_{\beta_0 \to 0} \beta_f^{\mathrm{ref}} &= 2 \lim_{r \ll 1} \lim_{\beta_0 \to 0} \beta_f^{\mathrm{abs}},
\end{align}

which states that a reflected photon imparts twice the momentum
as an absorbed photon (which can be seen to not hold in the
relativistic regime).

Consider a sail that is accelerated to relativistic speeds exclusively by 
absorbed photons, but maintained a constant temperature via
thermal equilibrium. This means that although the mirror's rest mass
temporarily increases after the absorption, it immediately re-radiates
this excess energy isotropically, thereby returning to a rest mass $m$.
Since isotropic re-radiation of the sail does not affect its velocity
(else anything moving and at non-zero temperature would feel a constant
drag/acceleration force), we may use the principle of ensemble equivalnce
used earlier in Section~\ref{sub:ensemble} to show that the velocity
curve is

\begin{align}
\beta_n^{\mathrm{abs}} &= \frac{n r}{1+nr}.
\label{eqn:betan_abs}
\end{align}

\subsection{Accounting for reflectivity}

We now need to combine the two cases, reflection and absorption, into a single
model described by a reflection coefficient, $R$. In what follows, we define
$R$ as being the fraction of incident photon power which is reflected
elastically by the mirror, with the remaining fraction $\bar{R}=(1-R)$ being
absorbed inelastically. For simplicity, we will also assume that the reflection
coefficient is achromatic.

We first point out that trying to derive this formula in the case of a single
photon appears intractable, on the basis that we have two conserved quantities
(energy and momentum) but three unknowns (final mirror velocity, final
frequency of the photon, and final rest mass of the mirror).

In order to make progress, we adopt the following approximate model.
We assume that a single photon can be split into two components, one of energy
$R h \nu_i$ which reflects off the sail, and the other of energy
$\bar{R} h \nu_i$ which is absorbed. Let's assume the elastic collision occurs
first, followed by the inelastic collision; in each independent collision, we
can analytically solve the final state of the system. In the time between this
``pair'' of photons and the next, we assume that the sail re-radiates the
excess absorbed energy, i.e. it is in thermal equilibrium. Using this model, we
can combine the results found earlier in Section~\ref{sub:single} \&
\ref{sub:inelastic} to write that, for an initial velocity of $\beta_0=0$, the
speed after the $n=1^{\mathrm{st}}$ incident photon is

\begin{align}
\beta_1^{\mathrm{mix}} &= 1 - \frac{1}{1 + r (1+R+2 R r)},
\end{align}

where the superscript ``mix'' on the left-hand side denotes that this
velocity change is a mixture model of both elastic and inelastic components.
Using our principle of ensemble equivalence (i.e. that a series of photon
impacts is equivalent to one cumulative energetic photon collision), we may
write that

\begin{align}
\beta_n^{\mathrm{mix}} &= 1 - \frac{1}{1 + n r (1+R+2 R n r)}.
\label{eqn:betamix}
\end{align}

As expected, Equation~(\ref{eqn:betamix}) can be easily demonstrated
to reproduce Equation~(\ref{eqn:betan}) in the limit of $R\to1$
and Equation~(\ref{eqn:betan_abs}) in the limit of $R\to0$.

\subsection{Numerical verification}

Equation~(\ref{eqn:betamix}) is derived by assuming that each photon can
treated as a pair of dummy photons. We test here the validity of
this assumption through numerical simulations.

In each simulation, we consider firing $10^5$ incident photons at a mirror
where the photon has a probability $R$ of being an elastic photon
and $\bar{R}$ of being inelastic\footnote{Practically speaking, we simply
generate a pseudo-random number between 0 and 1 and compare it to these
probabilities at each iteration}. Starting from rest, we numerically
compute the velocity curve of the mirror using Equation~(\ref{eqn:betaf})
for elastic collisions and Equation~(\ref{eqn:betaf_abs}) for those
which are inelastic. After each inelastic collision, we assume the mirror
re-radiates the absorbed energy isotropically before the next photon
arrives (i.e. thermal equilibrium), such that the rest mass of the mirror
does not evolve.

Since the simulations are intrinsically stochastic via the reflection 
probabilities, we repeat each simulation 1000 times and take the mean.
Because we have assumed a small number of incident photons (just $10^5$), we
use several large choices of $r=10^{-6}$, $10^{-5}$ \& $10^{-4}$ in order to
accelerate the mirror to relativistic speeds. We set the reflectivity to
$R=0.9$, representing a fairly poorly optimized sail. Comparing to the
predictions of Equation~(\ref{eqn:betamix}), we estimate the expression
is accurate to within 0.04\% for all reflectivities $R>0.9$.

\subsection{Velocity losses due to non-unity reflectivities}

We may now compare the velocity curve predicted by Equation~(\ref{eqn:betamix})
to that of a perfect mirror in Equation~(\ref{eqn:betan}), in order to quantify
the losses due to non-unity reflectivities:

\begin{align}
\Big(\frac{ \beta_n^{\mathrm{ref}} - \beta_n^{\mathrm{mix}} }{ \beta_n^{\mathrm{ref}} }\Big) &=
\frac{(1+2nr)(1-R)}{2(1+nr)(1+nr(1+R+2nrR)) },
\end{align}

which reduces to the classical result of 

\begin{align}
\lim_{r\to0}\Big(\frac{ \beta_n^{\mathrm{ref}} - \beta_n^{\mathrm{mix}} }{ \beta_n^{\mathrm{ref}} }\Big) &=
\Big(\frac{1-R}{2}\Big).
\end{align}

%

Since $n r$ dictates the final velocity of the mirror, we may replace $n r$
with the target velocity, $\beta_{\mathrm{targ}}$, and Taylor expand to first order in $\bar{R}$ to yield

\begin{align}
\Big(\frac{ \beta_n^{\mathrm{ref}} - \beta_n^{\mathrm{mix}} }{ \beta_n^{\mathrm{ref}} }\Big) &\simeq
\Bigg(\frac{(1-\beta_{\mathrm{targ}})\sqrt{1-\beta_{\mathrm{targ}}^2}}{(1 - \beta_{\mathrm{targ}}) + \sqrt{1-\beta_{\mathrm{targ}}}}\Bigg) \bar{R}.
\end{align}

Using the cumulative energy needed to accelerate to a perfect sail to
$0.2c$, the final speed of the sail is reduced by 4.4\% for a 90\%
reflectivity and 0.044\% for a 99.9\% reflectivity. We therefore conclude that
velocity losses due to imperfect reflectivities are fairly modest and
unlikely to be a limiting design constraint on the sail.

\subsection{Energy and thermal requirements}
\label{sec:thermal}

We also consider here the energy which is absorbed by the sail thermally.
One may re-arrange Equation~(\ref{eqn:betamix}) to write that the
cumulative number of photons needed to accelerate a sail up to a target
velocity $\beta_{\mathrm{targ}}$ is given by

\begin{align}
r_{\mathrm{tot}}(\beta_{\mathrm{targ}},R) &= \frac{1}{4 R (1-\beta_{\mathrm{targ}})} \Bigg( - (1+R)(1-\beta_{\mathrm{targ}}) \nonumber\\
\qquad& + \sqrt{ 8\beta_{\mathrm{targ}}R(1-\beta_{\mathrm{targ}})+(1-\beta_{\mathrm{targ}})^2(1+R)^2}  \Bigg),
\end{align}

where $r_{\mathrm{tot}} = n r$. The above can also be expressed as an
energy given by

\begin{align}
E_{\mathrm{inc}} =& m c^2 r_{\mathrm{tot}}(\beta_{\mathrm{targ}},R).
\end{align}

\begin{figure*}
\begin{center}
\includegraphics[width=16.0cm,angle=0,clip=true]{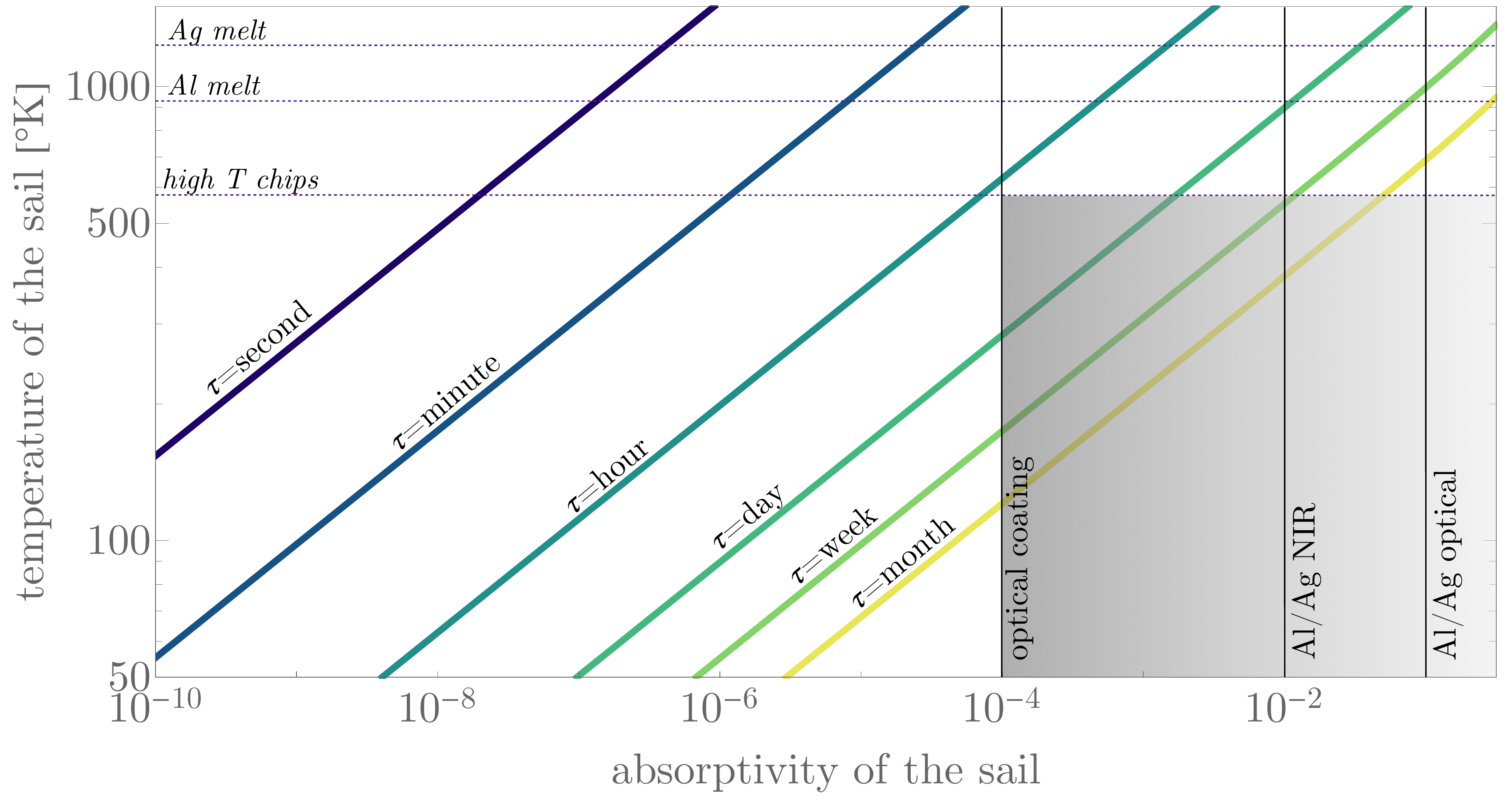}
\caption{\emph{
Predicted temperature of a 1\,g mass, 16\,m$^2$ relativistic sail
accelerated up to $0.2c$. Each contour shows the equilibrium temperature
of the sail as a function of its reflectivity for distinct laser firing
times. Firing for longer allows the sail more time to re-radiate the
absorbed thermal photons, but ultimately poses significant practical
challenges. The gray region denotes a physically plausible region of
parameter space, although future advancements could increase this region.
}}
\label{fig:thermal}
\end{center}
\end{figure*}

Note that this is the energy incident upon the sail and does not account for
beam losses due to diffraction or scattering between the laser source and the
sail. In total, we assume that the sail has absorbed a fraction $\bar{R}$ of
this energy as thermal photons over a time $t$. Time $t$ corresponds to the
time that the photon is actually received, not when it is emitted, due to
light travel time. Time $t$ will always exceed the emission duration but
the ratio is extremely close to unity at the start of the acceleration
(e.g. see Equation~\ref{eqn:timelag}), leading to the greatest thermal stress
on the sail. Since this regime sets the design constraints on the sail, the
time lag is unimportant for this purpose. We also highlight that for sails with
finite transmittance, the prescription given here can be easily modified by
attenuating the incident energy accordingly.

In the sail's reference frame, the incident energy is received over a
dilated of time $t'$. Assuming a constant acceleration (or force) applied to
the sail initially at rest, the time dilation factor is \citep{iorio:2005}

\begin{align}
\frac{t'}{t} &= \frac{\sinh^{-1}(\beta_{\mathrm{targ}})}{\beta_{\mathrm{targ}}}.
\end{align}

For $\beta_{\mathrm{targ}}=0.2$, this time dilation factor is less
than a percent and thus practically speaking one may simply assume
$t' \simeq t$.

As was done earlier, we assume that the sail immediately re-radiates the
absorbed energy isotropically. For the sake of simplicity, we assume that the
sail emits this thermal energy as a blackbody over the laser firing time of
$t$, such that the total energy emitted by the sail is $2 \mathcal{A} t
\sigma_B T^4$, where $\mathcal{A}$ is the area of the sail. Note that the
sail's area is not length contracted since it is normal to the direction of
motion. Equating the received and emitted powers and then solving for the
sail's temperature, $T$, we have

\begin{align}
T^4 =& \Big(\frac{\Sigma \bar{R} c^2}{2\sigma_B t}\Big) r_{\mathrm{tot}}(\beta_{\mathrm{targ}},R),
\label{eqn:temp}
\end{align}

where $\Sigma$ is the effective\footnote{We use the term ``effective'' because
the rest mass includes the payload} surface density of the sail, given by
$\Sigma \equiv m/\mathcal{A}$. Note that Equation~(\ref{eqn:temp})
refers to the temperature of the sail in the Earth's frame of reference,
not in the sail's frame of reference which we ultimately require.
\citet{einstein:1907} and \citet{planck:1908} argue that temperature is
covariant, given by $T'=T/\gamma$ (where $\gamma=1/\sqrt{1-\beta^2}$), but
\citet{ott:1963} later challenged this, obtaining the result $T'=T \gamma$.
Later, \citet{landsberg:1966,landsberg:1967} argue that thermodynamic
quantities like entropy and temperature should not vary between two reference
frames and we adopt this result in our work here too\footnote{These disagreements
provide an interesting opportunity for experiment onboard Starshot}, i.e. $T'=T$.

In order to proceed, we assign some parameters appropriate for the
\textit{Starshot} proposal. We choose optimistic but plausible values for the
spacecraft mass of $m=1$\,g and a sail of area $\mathcal{A}=16$\,m$^2$ and
assume $\beta_{\mathrm{targ}}=0.2$. The firing time is varied between
several options. We plot the resulting temperature of the
spacecraft, given by Equation~(\ref{eqn:temp}), as a function of aborptivity
in Figure~\ref{fig:thermal}. As noted earlier, these temperatures should be
treated as the temperature which the sail rises to during the initial
phases of acceleration, but finite light travel time will lead to a cooling
effect at later times.

As an example, for $\bar{R}=10^{-5}$ aborptivity, which is plausible with
optically coated materials \citep{rempe:1992}, temperatures below
$300^{\circ}$C (typical of a high-temperature microsystem;
\citealt{lien:2011,chiamori:2014}) can be maintained over an 8.6\,minute
firing period.

Such a case would require just over 10\,TJ of incident energy on the
sail, or a constant power of 19.6\,GW, giving an average flux on the sail of
1.2\,GW\,m$^{-2}$ for the adopted 16\,m$^2$ area.

\subsection{Diffraction Losses}

The rapid acceleration of the sail causes it to quickly traverse great
distances which poses at least two challenges for the laser system.
First, at great distances it may be difficult to maintain accurate
pointing on the sail, particularly if atmospheric turbulence introduces
small refractive deviations to the optical path. Second, even if perfect
pointing is maintained, diffraction of the laser light can introduce
significant losses of the beam energy by the time it reaches the target.
We tackle this second issue in what follows and assume a stable sail
riding the beam throughout (see \citealt{manchester:2017} for details
on this point).

Consider a transmitter of diameter $D_T$ producing a laser of wavelength
$\lambda$, which strikes its target at a distance of $L$ away from the source.
For a diffraction-limited beam, the beam width at distance $L$ will be
\citep{teachey:2016}

\begin{align}
W_L &= \frac{\sqrt{2}L\lambda}{D_T},
\end{align}

where we have assumed that the final beam width has diffracted to be much
greater than the initial width.

For simplicity, we consider a sail which is circular in projection and
a Gaussian beam profile. At a distance $L$ then, the integrated fraction
of the laser power striking the sail will be

\begin{align}
F_P &= \Bigg(\mathrm{erf}\Big[\frac{1}{W_L}\sqrt{\frac{\mathcal{A}}{2\pi}}\Big]\Bigg)^2.
\end{align}

The above can now be evaluated by replacing $L$ with the corresponding
distance expected at some target velocity, $\beta_{\mathrm{targ}}$. To
make analytic progress, we will assume that the sail undergoes strictly
uniform but relativistic acceleration. Accordingly, the distance the
sail has traversed

\begin{align}
L &= \frac{a_{\mathrm{const.}} t^2}{1 + \sqrt{1-\beta_{\mathrm{targ}}^2}},
\end{align}

where $a_{\mathrm{const.}}$ is the constant acceleration of the sail as
observed in the laser's reference frame, given by $c \beta_{\mathrm{targ}}/
t$. Using the above, the fractional power striking the sail is now

\begin{align}
\mathrm{erf}^{-1}\sqrt{F_P} &= \sqrt{\frac{\mathcal{A}}{4\pi}} \frac{D_T}{c} \frac{1}{\lambda t} \frac{1 + \sqrt{1-\beta_{\mathrm{targ}}^2}}{\beta_{\mathrm{targ}}}.
\end{align}

\begin{figure}
\begin{center}
\includegraphics[width=8.4cm,angle=0,clip=true]{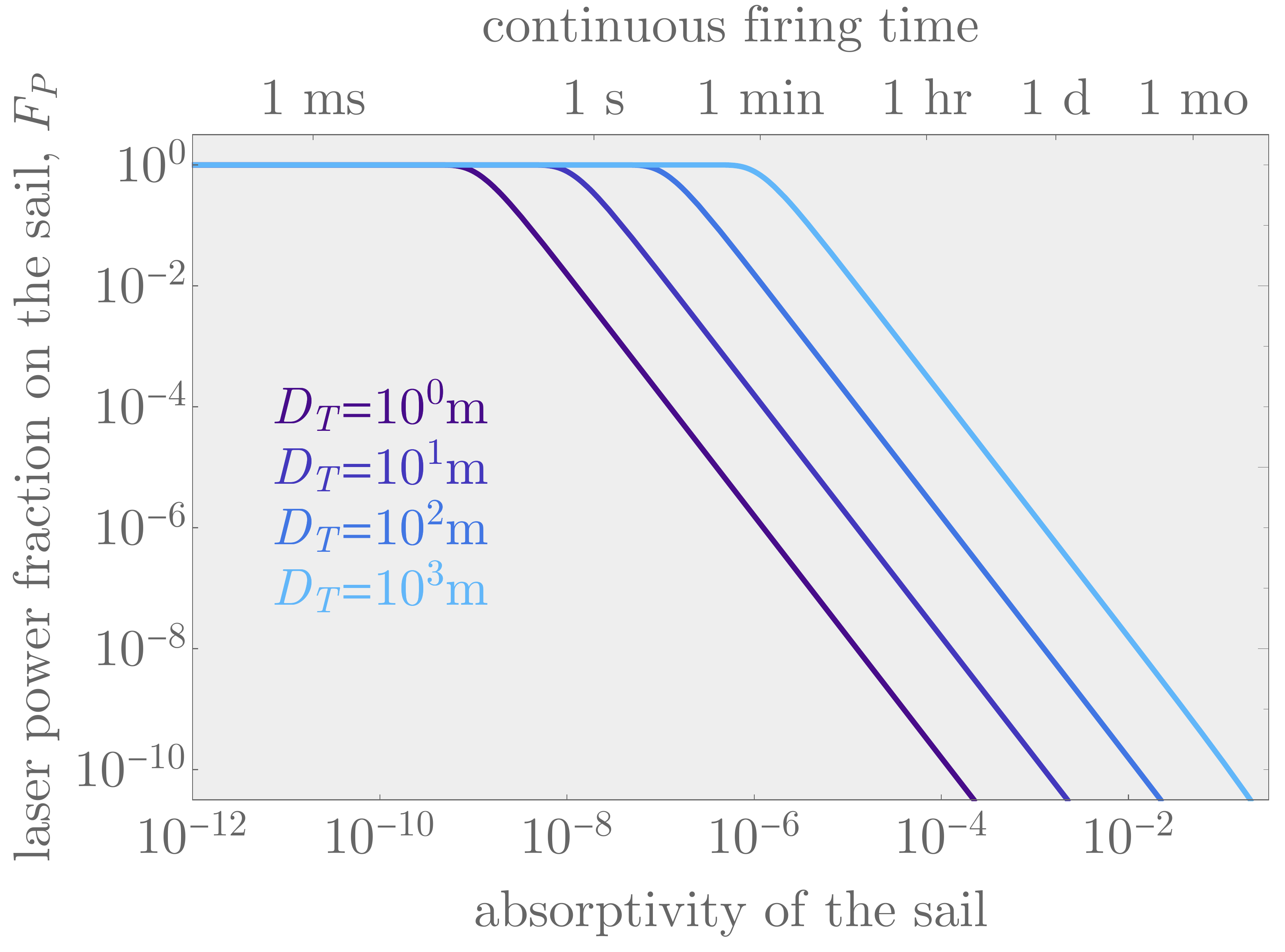}
\caption{
\emph{
Fraction of laser power which strikes the sail by the time that the
sail reaches 0.2c, as a function of firing times (top-axes). Since
firing time directly relates to the temperature of the
sail for a given reflectivity, the bottom axes depicts the corresponding
reflectivity of the sail in order to maintain temperatures below $300^{\circ}$C.
}
}
\label{fig:losses}
\end{center}
\end{figure}

One may now replace $t$ in the above with Equation~(\ref{eqn:temp}) to
relate the power loss at a given target velocity as function of the basic
sail properties. Before doing so, we present a quick an order-of-magnitude
calculation by using $L \simeq \tfrac{1}{2} \beta_{\mathrm{targ}} c t$
(non-relativistic) to give $L \simeq 2.9$\,AU. For a $D_T=10$\,m transmitter
at 650\,nm wavelength, the beam width will be $W_L\simeq40$\,km at a distance
$L$ and thus we should expect $(16/40000)^2 \sim 10^{-7}$ fractional power
striking the sail. Using the full equation, we obtain similar results, as
depicted in Figure~\ref{fig:losses} for four possible choices of $D_T$.

Our results imply that any firing time of order hours or greater will
lead to very large beam losses of at least a million, increasing the
energy demands to ten exa-joules or more. To keep energy losses within
a factor of 10, a kilometer sized transmitter could fire for 199\,seconds
at a sail with an absorptivity satisfying $\bar{R}<3.9 \times 10^{-6}$.
These calculations argue that key technical requirements for \textit{Starshot}
are a $\bar{R}=10^{-6}$ sail, kilometer-sized lasers achieving 500\,GW power
over a firing time of a few minutes.

\section{Discussion}
\label{sec:discussion}

We have derived an equation for the velocity change of a relativistic moving
mirror (or equivalently a light sail) in response to an ensemble of normal
incident photons, as well as the corresponding redshift of the reflected
photons. Whilst our formulae for the cases of a perfectly reflective and
perfectly absorptive mirror are exact, our formula for a mirror
with a reflectivity in the range $0<R<1$ should be treated as an excellent
approximation rather than being formally true, and we suggest that the
solution may in fact be intractable without approximation.

Crucially, our expression for the velocity curve differs from
that stated in \citet{lubin:2016}, which motivates the \textit{Starshot}
project \citep{popkin:2017}. The \citet{lubin:2016} result is derived
in \citet{kulkarni:2016}, who use Lorentz frame transfers and assume
that in the sail's rest frame the frequency of reflection equals that of
incidence. We have shown that this assumption, also made by Einstein in
his seminal 1905 paper introducing relativity, violates the conservation
of energy since the sail must increase it's momentum (and thus kinetic
energy) in response to the reflection and thus the photons must lose
energy by becoming redshifted. Since this treatement overestimates the
photon's final energy, it also underestimates the sail's velocity.
For \textit{Starshot}-like parameters, the difference is small, corresponding
to a difference of $\sim 50$m/s in the predicted speed of the sail.
If the arrival time of \textit{Starshot} needs to be predicted
to a precision of minutes, or equivalently the arrival position to
within a few Earth radii, then our formula should be favored over the
formalism of \citet{lubin:2016} and \citet{kulkarni:2016}.

Our equations provide an analytic framework to predict the
acceleration of a light sail under solar or laser irradiation up to
relativistic speeds, as appropriate for the \textit{Starshot} project
for example.

A useful result from our work is that the relativistic velocity curve from a
large number of incident photons can be described analytically as that of a
single photon with the equivalent energy of the ensemble, for either elastic
or inelastic collisions. This insight, which we have used several times and
referred to as the principle for ensemble equivalance for convenience, is
demonstrated in Section~\ref{sub:ensemble} and provides a simple analytic
approach for modeling sail response functions.

Additionally, we have discussed how the high levels of incident radiation on
the sail, necessary to achieve relativistic speeds, will put thermal stress on
the sail and payload. Practically speaking, the ideal sail material should be
ultra-light, rigid against the radiation pressure inhomogeneities,
thermally stable up to hundreds of Kelvin and ultra-reflective. To avoid
losing more than a factor of ten of the laser power through diffraction losses,
we find that a kilometer-sized transmitter needs to fire for 3.3\,minutes or
less, excerbating the thermal stress on the sail. For such a case, we estimate
that the absorptivity needs to less than $4 \times 10^{-6}$ and be able to
operate at $300^{\circ}$\,C (573 K).

There are numerous effects we have ignored which will further influence the
design requirements for \textit{Starshot}. For example, additional
beam losses due to scattering through the Earth's atmosphere will certainly
lead to a higher laser power output requirement than that estimated here. In
terms of the sail itself, at least three effects we have ignored will influence
the sail's velocity. First, drag forces from interstellar dust and even
photonic gas \citep{balasanyan:2009} will act to slowly decelerate the sail.
Second, we have assumed that the reflection coefficient is achromatic, but in
reality, man-made highly reflective surfaces, such as dielectric mirrors, are
often extremely sensitive to wavelength \citep{rempe:1992}. Third, we have
assumed that the sail and spacecraft chassis are in thermal equilibrium from
the first incident photon to the last, whereas in reality some of this energy
will not be re-radiated but used to warm up the chassis, potentially leading
to material deformations, for example. We highlight these problems to the
community for future work.

\acknowledgments

D.M.K. thanks members of the Cool Worlds Lab for stimulating conversations
on this topic. Thanks to Zoltan Haiman, Jules Halpern \& Emily Sandford for
their helpful comments on early drafts. Special thanks to the anonymous
reviewer for encouraging detailed literature comparisons.

\end{document}